# A general correlation for turbulent friction factors in non-Newtonian fluids

Trinh, Khanh Tuoc
Institute of Food Nutrition and Human Health
Massey University, New Zealand
*K.T.Trinh@massey.ac.nz*

---

## Abstract

Correlations of friction factors are presented for the general case of purely viscous non-Newtonian fluids without requiring a priori the adoption of a rheological model. They are based on an empirical estimate of the shift in the wall layer edge and the Kolmogorov point. The predictions of friction factors have the same level of accuracy as those of the Dodge-Metzner correlation but the visualisation is more compatible with measured velocity profiles. The general correlations obtained can be used to easily retrieve correlations for specific rheological models.

## 1 Introduction

There are a number of different generic approaches to the correlation of friction factors in non-Newtonian fluids. Firstly, the friction factor Reynolds number relationship may be expressed in terms of a logarithmic or a power law. Both of these have been shown have the potential for adequate predictions (Trinh, 2010a, Trinh, 2010b). Secondly we may attempt to derive formulae for specific rheological models the most popular being the Ostwald de Waele power law, the Bingham plastic and the Herschel-Bulkley models or obtain a relationship applicable to all non-Newtonian fluids without assuming a priori a fluid rheological model. A majority of published works deal with the Ostwald de Waele power law and have been reviewed in some detail in a previous paper (Trinh, 2009a). The Bingham plastic model has been used in works by Tomita (1959, 1985), Torrance (1963), Michiyoshi et al. (1965), Hanks and co-workers (Hanks and Dadia, 1971, Hanks and Valia, 1971), Kawase and Moo-Young (1992), Wilson and Thomas, (1985, 2006), The Herschel-Bulkley model has been used by Chilton and Stainsby (1998) and the Sisko model by Turian et al.(1998).

.Dodge & Metzner (1959) used dimensional analysis to extend Millikan's logarithmic law to time independent non-Newtonian fluids. The term time-independent means that the fluids do not thin (thixotropy) or thicken (rheopexy) with duration of shear and do not exhibit viscoelasticity. While Dodge and Metzner showed that a certain amount of drag reduction exists for time independent non-Newtonian fluids compared with Newtonian fluids at the same Reynolds drag reduction in the presence of viscoelasticity is much more pronounced as identified by Toms (1949) a distinction not made by their contemporary Shaver and Merrill (1959). A second distinctive feature of the Dodge and Metzner correlations is that they are expressed in terms of the behaviour index $n'$, the slope of the log-log plot between the shear stress τ and the flow function $(\Gamma = 8V/D)$ where V is the average flow velocity and D the pipe diameter and not the index n which is the slope of the log-log plot between $\tau_w$ and the shear rate $\dot{\gamma}_w$. Therefore the correlations are not restricted only to power law fluids obeying the correlation $\tau = K\dot{\gamma}^n$ but can be applied to time-independent fluids following any rheological model.

$$1/\sqrt{f} = 4.0/(n')^{0.75}\log\left[\mathrm{Re}_{MR}\, f^{1-n'/2}\right] - 0.4/(n')^{1.2} \qquad (1)$$

Where $\mathrm{Re}_g$ is the so called Metzner-Reed generalised Reynolds (MRR) number

$$Re_g = \frac{D^{n'} V^{2-n'} \rho}{K' 8^{n'-1}} \qquad (2)$$

Dodge and Metzner also proposed an empirical extension to the Blasius (1913) power law correlation.

$$f = a/\mathrm{Re}_{MR}^{b} \qquad (3)$$

Where

$$\alpha = 0.0665 + 0.01175n' \qquad (4)$$

$$\beta = 0.365 - 0.177n' + 0.062n'^2 \tag{5}$$

The Dodge and Metzner correlations have been widely accepted from the moment they were published and are routinely quoted in books on non-Newtonian fluid technology e.g. (Skelland, 1967, Chabra and Richardson, 1999, Steffe, 1996) and remain highly recommended even in recent evaluations of correlations for friction factors in power law fluids e.g. Gao and Zhang (2007).

However, the predictions of velocity profiles proposed by Dodge & Metzner with a slope $A = 2.46 / n'^{0.75}$, based on the success of their friction factor correlation, did not agree with the subsequent measurements of Bogue and Metzner (1963) where the slope of the log-law is described by $A = 2.5 / n'$.

In this paper we explore the possibility of developing a general correlation for turbulent pipe flow of all purely viscous non-Newtonian fluids that can then be easily adapted to any specific rheological model.

# 2 Theoretical considerations

## 2.1 Logarithmic correlations

The log-law can be written as

$$U^+ = \frac{1}{\kappa} \ln y^+ + B \tag{6}$$

where the velocity and distance $U^+ = U/u_*$ and $y^+ = yu_*\rho/\mu$ have been normalised with the friction velocity $u_* = \sqrt{\tau_w/\rho}$ and the fluid apparent viscosity $\mu$.

There are two major difficulties in estimating $U_\nu^+$, $\delta_\nu^+$ in the general case. First is a proper definition of the viscosity coefficient. The Mooney-Rabinowitsch equation for laminar pipe flow can be written as (Skelland, 1967)

$$\dot{\gamma}_w = \frac{8V}{D}\left(\frac{3n'+1}{4n'}\right) \tag{7}$$

The wall shear stress is

$$\tau_w = K'\left(\frac{8V}{D}\right)^{n'} \tag{8}$$

The non-Newtonian apparent viscosity at the wall is thus

$$\mu_w = \frac{\tau_w}{\dot{\gamma}_w} = \frac{K'(8V/D)^{n'-1}}{((3n'+1)/4n')} \tag{9}$$

It is used to define a non-Newtonian Reynolds number.

$$\mathrm{Re}_w = \frac{DV\rho}{\mu_w} \tag{10}$$

which has been used for example by Edwards (1980). By rearranging equation (10) we obtain a family of lines for the friction factor-Reynolds number plot in laminar pipe flow

$$f = \frac{16}{\mathrm{Re}_w}\left(\frac{4n'}{3n'+1}\right) \tag{11}$$

An effective viscosity is defined by Metzner and Reed (1955)

$$\mu_e = K'\left(\frac{8V}{D}\right)^{n'-1} \tag{12}$$

to collapse all friction factor-Reynolds number plots in laminar flow.

$$f = \frac{16}{\mathrm{Re}_g} \tag{13}$$

but it has no physical significance in turbulent pipe flow (Trinh, 1969). Nevertheless the MRR number has been widely used because of the success of the Dodge-Metzner correlation.. The distance in velocity profiles has been commonly normalised with the wall shear velocity and the apparent viscosity. Since an exact viscosity term in turbulent flows was not available these velocity profiles had to be represented by adopting a rheological model for the fluid. Bogue and Metzner for example used the power law model and obtained

$$y^+ = \frac{yu_*\rho}{\mu_w} = \frac{yu_*\rho}{K\tau_w^{\frac{n-1}{n}}} = \frac{yu_*^{2-n}\rho}{K} \tag{14}$$

The apparent viscosity in equation (16) is based on the time averaged wall shear stress but since the ejections of wall fluids create shear layers outside the wall layer depend on local instantaneous conditions we have to ask whether a more appropriate apparent viscosity should be based on the local shear stress which is easily shown from a force balance to vary linearly with radial distance

$$\tau = \tau_w \left( 1 - \frac{y}{R} \right) \tag{15}$$

This would significantly alter the value of $\kappa$ in measured velocity profiles. And then there is the proof in a previous paper (Trinh, 2009a) that the thickness of the wall layer is the same for Newtonian and non-Newtonian power law fluids when normalised with the instantaneous wall shear stress at the point of bursting. Would we learn more by using this instantaneous wall shear stress in the representation of velocity profiles? Unfortunately there is no measurement of the instantaneous wall shear stress in experimental studies of velocity profiles as far as I know.

Discussions about a correct definition of viscosity in non-Newtonian studies are not idle speculations for scientific interest, they are linked to visualisations of the physical mechanisms in turbulent flows. By normalising the velocity and distance with parameters at the edge of the wall layer $U/U_v$ and $y/\delta_v$ in the zonal similarity analysis (Trinh, 2010d), we sidestep the effect of variable viscosity on the velocity profile and can confirm that the value $\kappa = 0.4$ has widespread applicability.

### 2.1.1 Use of wall layer thickness

To define the log-law completely we have to estimate of the velocity and distance scale at the interface between the wall and log-law layers $U_v^+$ , $\delta_v^+$ .

$$B = U_v^+ - 2.5 \ln \delta_v^+ \tag{16}$$

This is the second major difficulty. An analytical solution for power law fluids has already been presented (Trinh, 2009a) and the resulting values of $U_v^+$ , $\delta_v^+$ have been used successful to obtain correlations for the friction factor (Trinh, 2010a, Trinh, 2010b). This can easily be redeveloped for other fluid models like the Bingham plastic or Herschel-Bulkley as shown in subsequent postings.

During the analysis of hundreds of velocity profiles to compile the zonal similarity master curve (Trinh, 2010d), it was found that the intersection of the Newtonian equation

$$\delta_\nu^+ = 4.16 U_\nu^+ \tag{17}$$

with experimental profiles in all fluids gave almost the same values of $U_\nu^+$ and $\delta_\nu^+$. as when separate correlations were used for each fluid model. For example, the relation for power fluids was found to be (Trinh, 2009a)

$$\delta_\nu^+ = 2.08\left(n+1\right)^{1/n} U_\nu^+ \tag{18}$$

While the velocity data for non-Newtonian fluids is scarce and not very accurate it is very clear that the $U_\nu^+$ and $\delta_\nu^+$ are always larger in non-Newtonian. Within the scatter of literature data used, a reasonable correlation was found to be

$$\delta_\nu^+ = 64.8\left(\frac{3n'+1}{4n'}\right) \tag{19}$$

The velocity at the edge of the wall layer becomes

$$U_\nu^+ \approx \frac{\delta_\nu^+}{4.16} = 15.58\left(\frac{3n'+1}{4n'}\right) \tag{20}$$

The generalised log-law for purely viscous non-Newtonian fluids can thus be written as

$$U^+ = 2.5\ln\frac{y^+}{64.8\left(\frac{3n'+1}{4n'}\right)} + 15.58\left(\frac{3n'+1}{4n'}\right) \tag{21}$$

$$\begin{aligned} U^+ &= 2.5\ln y^+ + 15.58\left(\frac{3n'+1}{4n'}\right) - 10.428 - 2.5\ln\left(\frac{3n'+1}{4n'}\right) \\ U^+ &= 2.5\ln y^+ + 1.44 + \frac{3.90}{n'} - 2.5\ln\left(\frac{3n'+1}{4n'}\right) \end{aligned} \tag{22}$$

Following the analyses of Prandtl (1935) and von Karman (1934) we ignore the relatively small effect of the law of the wake in closed channels and apply equation (22) up to $y^+ = R^+$. We further introduce the average velocity which is related to the maximum velocity by Prandtl as

$$U_m^+ = 4.03 + V^+ \tag{23}$$

Equation (23) only applies to Newtonian fluids and the constant $4.03$ will depend on $n'$ in the general case but since the wall layer is thin compared with the log-law layer in turbulent flows we will keep applying equation (23) as a simple good approximation. Equation (22) becomes

$$V^{+}=\sqrt{\frac{2}{f}}=2.5\ln R^{+}+15.58\left(\frac{3n'+1}{4n'}\right)-14.458-2.5\ln\left(\frac{3n'+1}{4n'}\right) \tag{24}$$

Substituting for

$$R^{+}=\mathrm{Re}_{g}^{1/n'}\, f^{\frac{1}{n'}-\frac{1}{2}}\left(\frac{3n'+1}{4n'}\right)2^{\frac{5n'-8}{2n'}} \tag{25}$$

$$Re_{g}=\frac{D^{n'}V^{2-n'}\rho}{K'\,8^{n'-1}} \tag{26}$$

$$\frac{1}{\sqrt{f}}=\frac{4.06}{n'}\log\left(\mathrm{Re}_{g}\, f^{1-\frac{n'}{2}}\right)+1.96-\frac{2.754}{n'} \tag{27}$$

### 2.1.2 Use of the Kolmogorov point

There is a more convenient technique for defining equation (6) completely. It consists in forcing it through its intersection with the line

$$U^{+}=y^{+} \tag{28}$$

Equation (28) was originally used by Prandtl (1935) to validate the existence of a so-called steady state laminar sub-layer that was to apply in Newtonian fluids to the range $0<y^{+}<5$. Modern visualisations of the wall layer process starting with Kline et al. (1967) have clearly shown that the there is no steady state laminar flow in the wall layer but the unsteady viscous diffusion of momentum in the wall layer can be well correlated (Einstein and Li, 1956, Meek and Baer, 1970, Trinh, 2009b) by time-averaging the solution of Stokes first problem (1851)

$$\frac{U}{U_{v}}=erf\left(\frac{y}{\sqrt{4\nu t}}\right) \tag{29}$$

This solution confirms that the time-averaged equation (29) does coincide with equation (28) in the range $0<y^{+}<5$ but departs from it significantly in the range $5<y^{+}<\delta_{v}^{+}$. For Newtonian fluids the intersection between equation (28) with the Prandtl-Nikuradse log law

$$U^{+}=2.5\ln y^{+}+5.5 \tag{30}$$

occurs at $U_k^+ = y_k^+ = 11.8$ that I have named the Kolmogorov point in respect for his elegant and influential contribution to turbulence research. It has been used by many researchers as an anchor point for derivations of turbulent transport correlations e.g. (Prandtl, 1910, Levich, 1962, Trinh, 1969, Wilson and Thomas, 1985). The Kolomogorov point does not coincide with any real coordinate on measured velocity profiles and does not represent the size of the smallest energy dissipating eddies as argued by some (Wilson and Thomas, 1985). In fact it occurs to the left of the wall layer velocity profile therefore the implied shear stress at the intersection is lower than the contribution of viscous shear alone which is physically unrealistic. However at the distance $y_k^+$ the so-called laminar and turbulent local shear stresses are equal

$$\tau_v = \mu \frac{dU}{dy} = \tau_t = \frac{\tau}{2} \tag{31}$$

and in that sense it corresponds to the Kolmogorov postulate (1941) that there is a scale defined by the equality between the viscous and turbulent energies

$$l_k = \frac{\mu^{3/4}}{\rho^{1/2}} \varsigma^{-1/4} \tag{32}$$

where $\zeta$ is the kinetic energy per unit volume. In analogy with equation (19) the Kolmogorov point may be estimated by (Trinh, 1969)

$$U_k^+ = y_k^+ = 11.8\left(\frac{3n'+1}{4n'}\right) \tag{33}$$

When equation (6) is forced through the Kolmogorov point, it becomes

$$U^+ = 2.5\ln y^+ + 2.68 - 2.5\ln\left(\frac{3n'+1}{4n'}\right) + \frac{2.95}{n'} \tag{34}$$

$$\frac{1}{\sqrt{f}} = 4.06\log(\mathrm{Re}_g f^{1-\frac{n'}{2}}) + 2.10 - \frac{2.81}{n'} \tag{35}$$

## 2.2 Blasius power law

A Blasius type equation can be obtained by forcing the $1/7th$ velocity profile through the Kolmogorov point. Using the same technique as before (Trinh, 2010b) gives

$$f = \frac{\alpha}{\mathrm{Re}_g^{(1/3n'+1)}} \tag{36}$$

$$\alpha = \frac{11.8^{-\frac{6n'}{3n'+1}} \left( \frac{3n'+1}{4n'} \right)^{\frac{7n'}{3n'+1}} 2^{\frac{4+n'}{3n'+1}}}{0.817^{\frac{7n'}{3n'+1}}} \tag{37}$$

Equation (9) shows that the factor $(3n'+1/4n')$ can be written as the ratio of the non-Newtonian and Newtonian shear rates for the same shear stress

$$\left( \frac{3n'+1}{4n'} \right) = \frac{\dot{\gamma}_{w,non-Newtonian}}{\dot{\gamma}_{w,Newtonian}} \tag{38}$$

and gives a simple way to apply the derivation to any rheological model. For example the apparent viscosity in Bingham plastic fluids

$$\tau = \tau_y + \eta\dot{\gamma} \tag{39}$$

where $\eta$ is the plastic viscosity and $\tau_y$ is the yield stress is

$$\mu_w = \frac{\eta\tau_w}{\tau_w - \tau_y} = \frac{\eta}{1-c} \tag{40}$$

$c = \tau_y / \tau_w$. The log law is then

$$U^+ = 2.5 \ln\left( \frac{yu_*\rho}{\eta}(1-c) \right) + B \tag{41}$$

The Buckingham equation (Skelland, 1967)

$$\frac{8V}{D} = \frac{\tau_w}{\eta}\left( 1 - \frac{4}{3}c + \frac{1}{3}c^4 \right) \tag{42}$$

Allows us to write for the Kolmogorov point

$$U_k^+ = y_k^+ = 11.8\left( \frac{1-c}{1-\frac{4}{3}c+\frac{1}{3}c^4} \right) \tag{43}$$

Then

$$U^+ = 2.5 \ln\left[ \frac{yu_*\rho}{\eta}\left( 1 - \frac{4}{3}c + \frac{1}{3}c^4 \right) \right] + 11.8 \frac{1-c}{1-\frac{4}{3}c+\frac{1}{3}c^4} - 6.17 \tag{44}$$

Using again equation (23) we obtain

$$\frac{1}{\sqrt{f}} = 4.07 \log\left[\frac{DV\rho}{\eta}\sqrt{f}\left(1-\frac{4}{3}c+\frac{1}{3}c^4\right)+8.35\frac{1-c}{1-\frac{4}{3}c+\frac{1}{3}c^4}-9.06\right] \qquad (45)$$

# 3 Results and discussion

The derivations presented here have been compared with the experimental data of Dodge (1959), Bogue (1962) and Yoo (1974) for the general fluid and the data of Thomas (1960) for Bingham plastic fluids.

Equation (35) is plotted in Figure 1. It correlated 270 data points with a standard deviation of 4.43% and a maximum deviation of 16.8%. Sixteen (16) predicted points were more than 5% above measured values and 48 points more than 5% below. Equation (23) is not shown because it is very close to equation (35). The Dodge-Metzner correlation applied for the same data yielded a standard deviation of 3.92% and a maximum deviation of 19.4%. The Dodge-Metzner correlation is plotted in red lines in Figure 1. Thus the same accuracy is obtained with equation (35) but it does not require a log-law slope of $A = 2.46 / n'^{0.75}$ which is not shown in the measurements of Bogue (1962) and Clapp (1961).

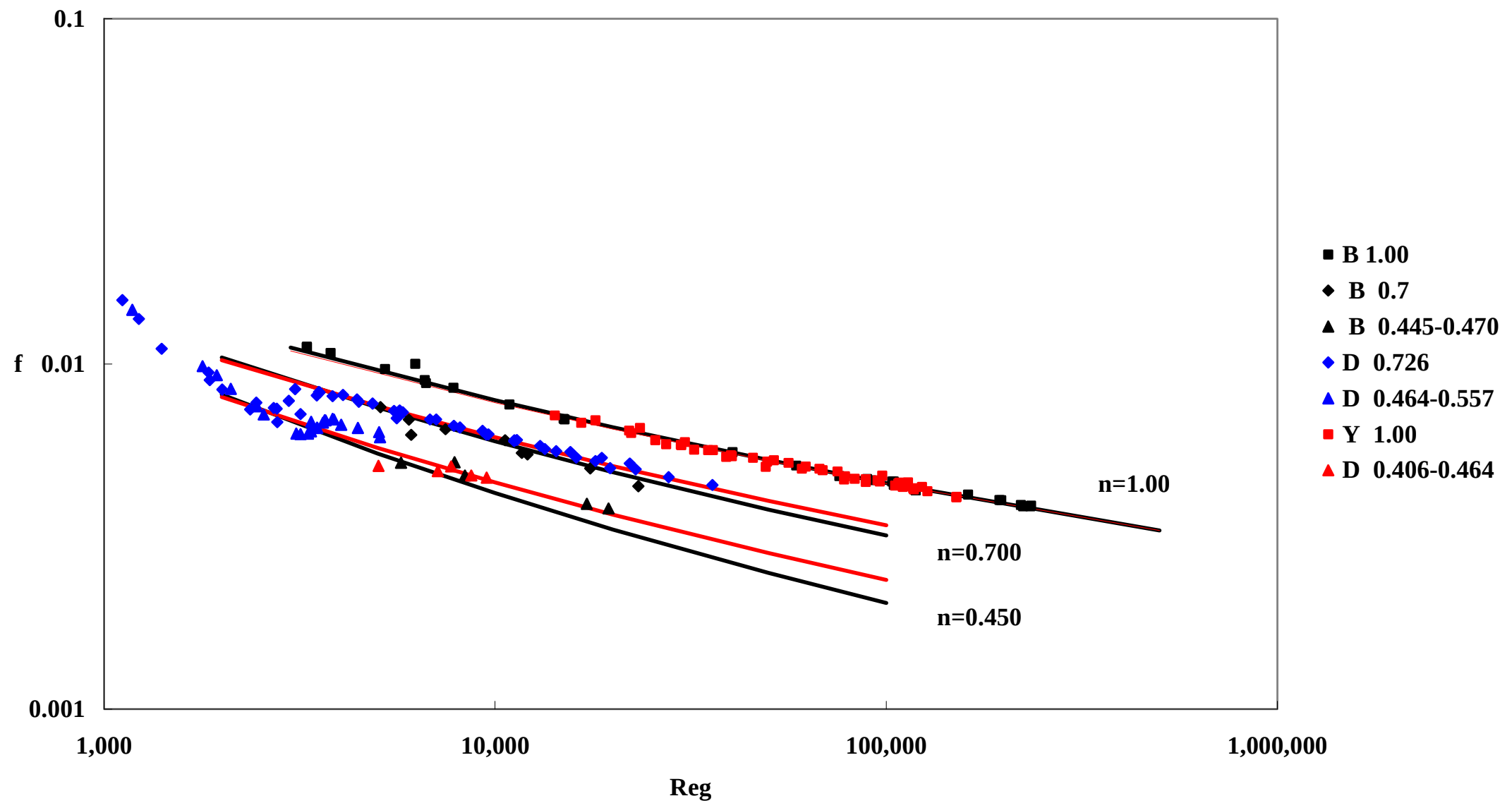

Figure 1. Correlations of friction factors (Red lines, equation (1) Dodge-Metzner, black lines, equation (35). Data of Dodge (1959), Bogue (1962) and Yoo (1974).

There is also a difference between the constant $0.4/n'^{1.2}$ in the Dodge-Meztner equation and curve fitting and the two terms constant $2.16-(2.78/n')$ in equation (35). The later is linked with physical visualisation of apparent thickening of the wall layer whereas the former is simply assumed for convenience as long as it complies with dimensional analysis. Equation (35) can be improved quite easily by taking account of $n'$ in equation (23), which strictly speaking only applies to Newtonian solutions. This modification has not been introduced here because such marginal improvement at the cost of simplicity of formula does not seem justified in view of the inherent limited accuracy in measurements in non-Newtonian flow. An example of such inaccuracy in the Dodge data has been given in a previous paper (Trinh, 2010c). Wilson and Thomas (1985) also used the principle of forcing the velocity profile through the Kolmogorov point but estimated it as $y_k^+ = 11.6(2/1+n)$. The factor $2/1+n$ arises from an integration of the shear stress-shear rate curve and applies to all flow geometries. Their predictions fall 5% to 15% below the predictions of Dodge and Metzner.

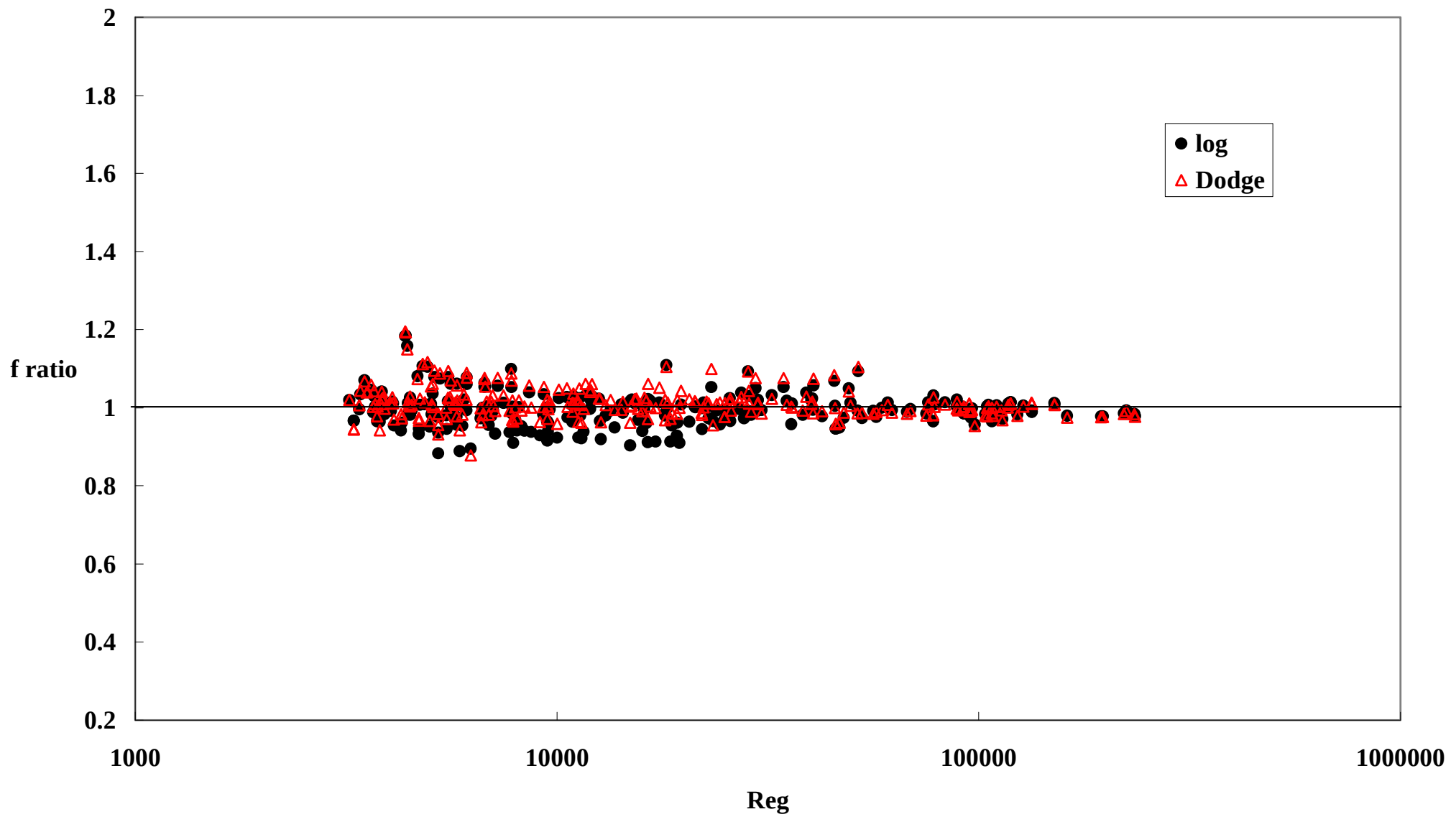

Figure 2 Comparison between the ratio of predicted to experimental friction factors between equations (1) and (35).

A comparison of the ratio of predicted to experimental friction factors by equation (1) and (350 is shown in Figure 2.

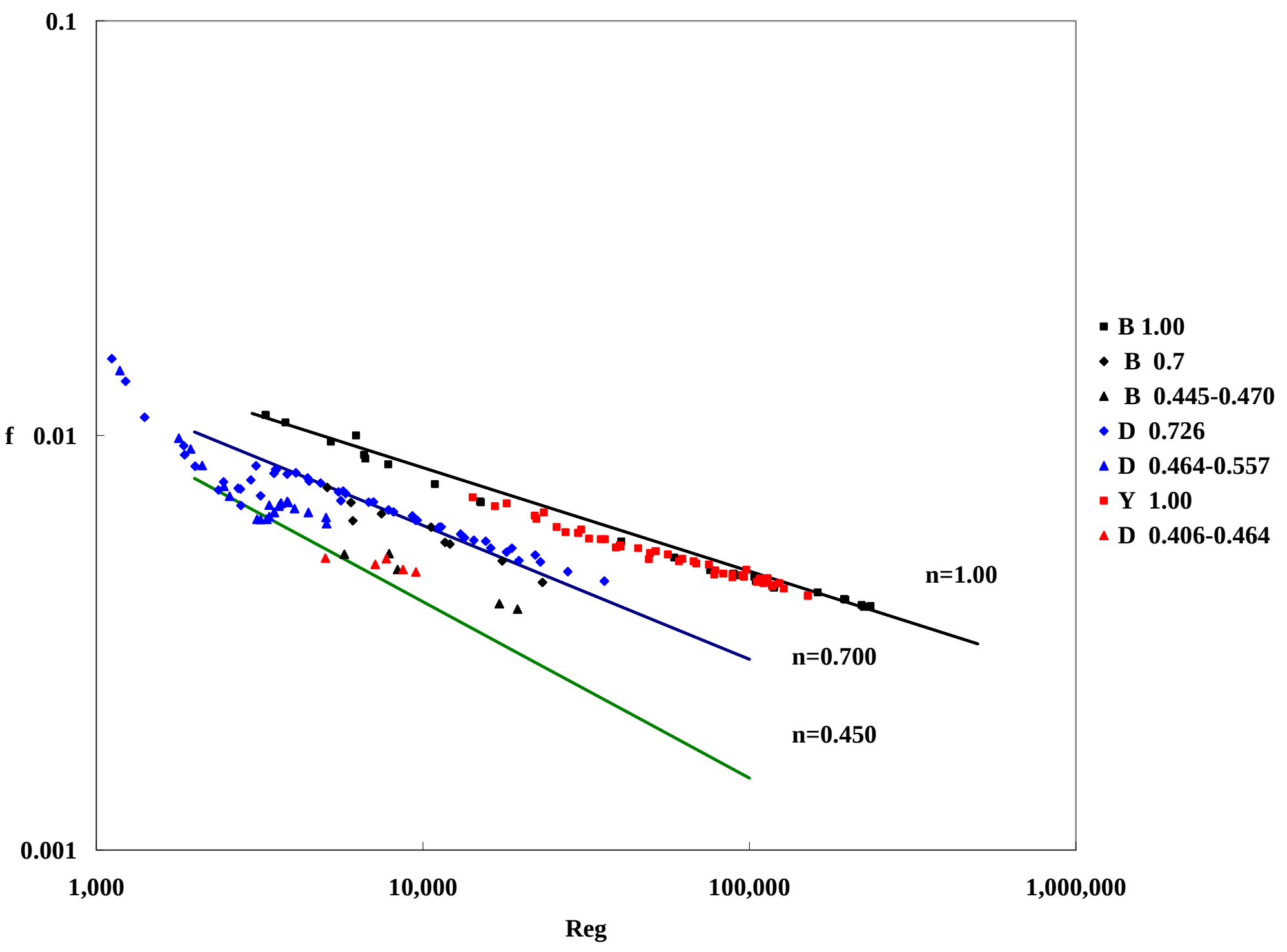

Figure 3 Correlations of friction factors with the power law equation (36).

Equation (36) is shown in Figure 3 against the same data. The standard deviation 6.49% and the maximum deviation 19.4%. The correlation is less accurate at higher Reynolds numbers and lower values of $n'$ because the $1/7th$ velocity profile does not apply well in these circumstances. It can be improved by a procedure used to develop an explicit correlation for external boundary layer flows (Trinh, 2009b).

The difference between equations (35) and (36) is seen more clearly in Figure 4 where the Blasius type correlation tends to over predict friction factors at higher Reynolds numbers.

An example of application to Bingham plastic fluids is shown in Figure 5. Equation (43) correctly predicts the measured friction factors at high Reynolds numbers but does not perform well in the intermediate regions right after the transition from laminar flow. This discrepancy occurs when the value for $c$ is relatively high and disappears at high Reynolds numbers because the wall shear stress is increased and $c \rightarrow 0$. In such cases equation (23) does not apply well because there is a plug in the region $0 < r_y < cR$ where the shear rate is zero and the velocity is constant.

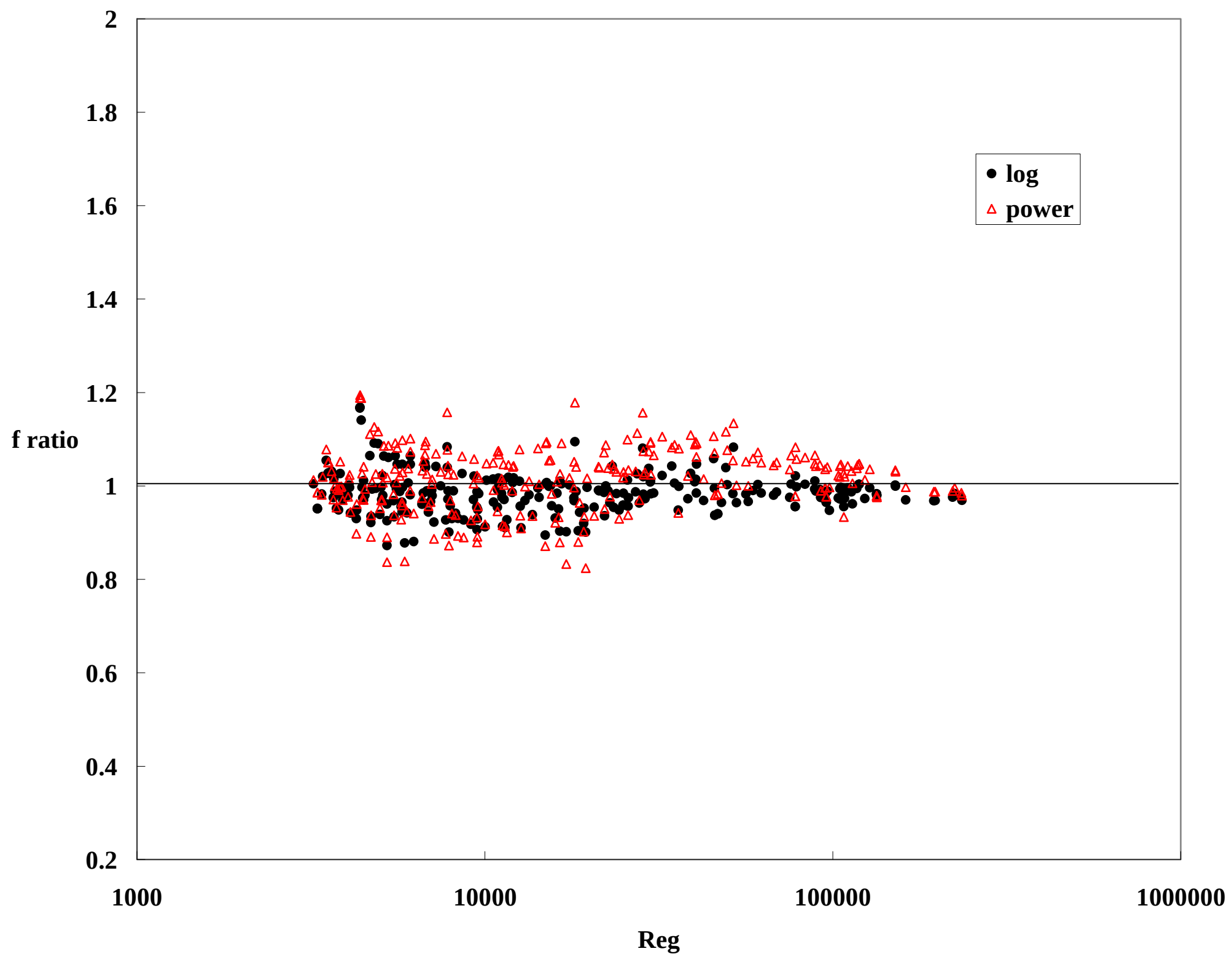

Figure 4 Comparison between the ratio of predicted to experimental friction factors between equations (35) and (36)

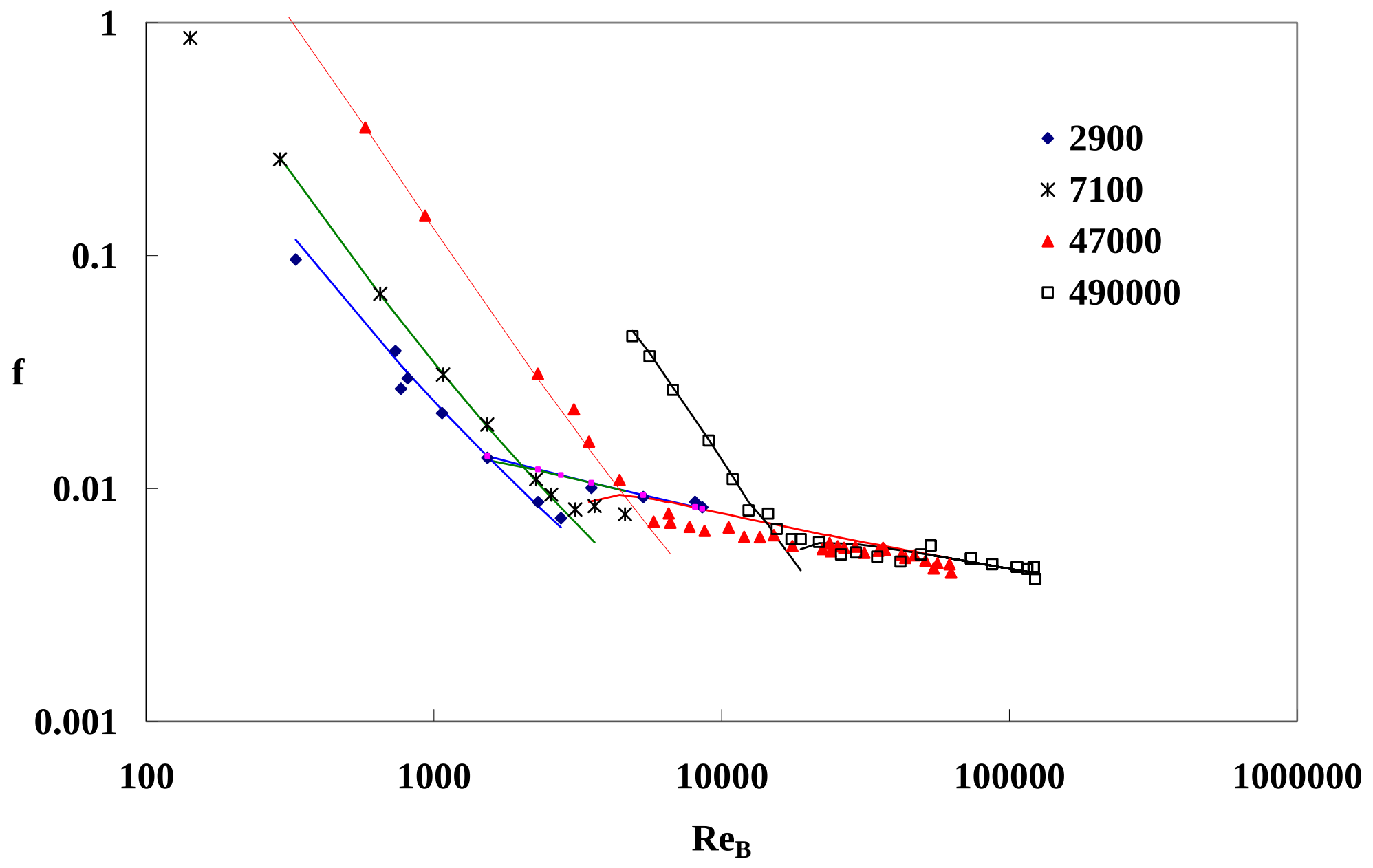

Figure 5 Plot of equation (43) against the data of Thomas (1960) for thorium suspensions. Numbers indicate the Hedstrom number.

A better correlation than equation (23) can be developed by numerical integration of the velocity profile. Improved correlations for Bingham plastic fluids will be discussed in another posting. An example for the Ellis model has been shown in Trinh (1969).

# 4 Conclusion

An approach to correlations of friction factors in purely viscous non-Newtonian fluids has been presented based on an empirical estimate of the shift in the wall layer edge and the Kolmogorov point. The predictions of friction factors have the same level of accuracy as those of the Dodge-Metzner correlation but the visualisation is more compatible with measured velocity profiles. The general correlations obtained can be used to easily retrieve correlations for specific rheological models. A simplified example is given for Bingham plastic fluids.